\title{Early Requirements Traceability with Domain-Specific Taxonomies - A Pilot
Experiment}

\documentclass[conference]{IEEEtran}

\usepackage[T1]{fontenc}
\usepackage[utf8]{inputenc}
\usepackage{booktabs}
\usepackage{tabularx}
\usepackage{graphicx}
\usepackage{todonotes}
\usepackage[bookmarks=false]{hyperref}
\usepackage{amsmath}
\usepackage{caption}
\usepackage[flushleft]{threeparttable}

\begin{document}

\author{\IEEEauthorblockN{Michael Unterkalmsteiner}
  \IEEEauthorblockA{Department of Software Engineering\\
Blekinge Institute of Technology\\
Karlskrona, Sweden\\
michael.unterkalmsteiner@bth.se}}

\maketitle

\begin{abstract}
  \emph{Background:} Establishing traceability from requirements documents to
  downstream artifacts early can be beneficial as it allows engineers to reason
  about requirements quality (e.g. completeness, consistency, redundancy).
  However, creating such early traces is difficult if
  downstream artifacts do not exist yet. \emph{Objective:} We propose to use
  domain-specific taxonomies to establish early traceability, raising the value
  and perceived benefits of trace links so that they are also available at later
  development phases, e.g. in design, testing or maintenance. 
  \emph{Method:} We developed a recommender system that suggests trace links
  from requirements to a domain-specific taxonomy based on a series of
  heuristics. We designed a controlled experiment to compare industry
  practitioners' efficiency, accuracy, consistency and confidence with and
  without support from the recommender. \emph{Results:} We have piloted the
  experimental material with seven practitioners. The analysis of self-reported
  confidence suggests that the trace task itself is very challenging as both
  control and treatment group report low confidence on correctness and
  completeness. \emph{Conclusions:} As a pilot, the experiment was successful since
  it provided initial feedback on the performance of the recommender, insight on
  the experimental material and illustrated that the collected data can be
  meaningfully analysed.
\end{abstract}

\begin{IEEEkeywords}
  Traceability, Requirements, Domain-specific Taxonomy, Recommender, Pilot Experiment 
\end{IEEEkeywords}

\section{Introduction}
Tracing requirements to downstream artifacts has benefits, such as more
efficient and correct software maintenance~\cite{mader_developers_2015}, enables
requirements-based testing~\cite{bouillon_survey_2013}, and is often demanded by
regulations on software production~\cite{regan_traceability-why_2012}. However,
requirements engineers lack motivation to create traces as they usually are not
the beneficiaries of traceability~\cite{arkley_overcoming_2005}. Furthermore,
creating traces from requirements to downstream artifacts requires that these
downstream artifacts already exist. Early requirements traces would increase
their value for engineers, as they would allow to reason about requirements
correctness, completeness or consistency.

To reap these early trace benefits for requirements engineers, we propose to
trace requirements to domain-specific taxonomies. This enables early
requirements analysis by exploiting the information encoded in a taxonomy, i.e.
the definitions and hierarchies of domain concepts. For example, engineers can
reason about the completeness and correctness of requirements
specifications~\cite{dzung_improvement_2009,kof_ontology_2010,moser_requirements_2011}.
This is particularly useful when the number of requirements
is high (in the order of thousands) and are written over a longer period of time
by different engineers. Furthermore, when information systems are used where
direct traces between artefacts are not possible (e.g. in an outsourcing
scenario), the taxonomy serves as an ``index'' to establish traceability. 

The approach to use a domain-specific taxonomy as a mean to enable early
traceability across time, infrastructure and organizational borders is contingent
on two assumptions: (1) such a taxonomy exists or can be created at low cost.
In this paper, we assume that such a taxonomy exists; (2) engineers are able
to associate taxonomy concepts to requirements. The evaluation of this second
assumption is subject of an ongoing research project. The main research question
we address is:
\begin{itemize}
\item[RQ] To what extent can a recommender system support practitioners in
  associating requirements with concepts defined in a taxonomy?
\end{itemize}

We implemented a recommender system, $CCR$, that we are currently evaluating
with practitioners. In this paper, we report on a pilot that aimed at validating
the experiment design, material and instrumentation. As the number of
participating subjects in the pilot was seven, we are not able yet to answer our
main research question. However, these initial results provide an indication
that the task, associating requirements with taxonomy objects, is difficult,
even for domain experts. Furthermore, we illustrate metrics which could be
helpful for researchers evaluating recommender systems in general.

The remainder of the paper is structured as follows. In Section~\ref{sec:brw} we
provide background on knowledge organization systems, introduce the basic
underpinnings of $CCR$ and point to related work. We document the experimental
plan in Section~\ref{sec:ep}, analyze the results in Section~\ref{sec:ana} and
discuss the outcome of the pilot experiment in Section~\ref{sec:dis}. We
conclude the paper in Section~\ref{sec:con}.

\section{Background and Related Work}\label{sec:brw}
\subsection{Knowledge Organization Systems}\label{sec:tax}
There exists a wide spectrum of processes and artifacts that are used in practice to
represent entities and their relationships for various knowledge oriented
applications~\cite{gruninger_ontology_2008,garshol_metadata_2004}. 
A controlled vocabulary is a closed list of terms that describes a subject area.
It homogenizes the assignment of terms to concepts as everyone is limited to the
same, existing definitions (for example, the Library of Congress Subject
Headings\footnote{\url{http://id.loc.gov/authorities/subjects.html}}).
Taxonomies provide additional structure to controlled vocabularies. The term
``taxonomy'' is rooted in the greek \emph{taxis}, which broadly means the
arrangement of things, and \emph{nomos}, meaning law or science. A taxonomy is a
set of rules to order things, abstract or concrete. Taxonomies can, but don't
have to, be hierarchical and encode a particular slice of knowledge about the
world. Modifications to content and structure typically require the consensus of
the community that has ownership over the taxonomy. The IEEE
Thesaurus\footnote{\url{https://www.ieee.org/publications/services/thesaurus.html}}
is an example that contains engineering, technical and scientific terms,
organized in a broad to narrow progression. While taxonomies are useful
constructs to encode and share knowledge, they are one-dimensional and based on
a closed vocabulary~\cite{garshol_metadata_2004}. Ontologies allow to model
relationships between concepts, even if they are part of different taxonomies.
This makes ontologies versatile in representing domain knowledge, requiring
however also formal languages, such as the Web Ontology Language
(OWL)\footnote{\url{https://www.w3.org/TR/owl2-overview/}}, that enable
computational reasoning and applications in artificial
intelligence~\cite{garshol_metadata_2004}. 

In the remainder of this paper, we focus on the use of taxonomies as knowledge
organization systems since we studied the proposed approach, described next, in
a context in which a domain-specific taxonomy already exists.
Nevertheless, we point out that the basic idea we propose is independent from the
underlying knowledge structure. The type of analyses, however, that can be performed
once requirements are traced, depends on the sophistication of the used knowledge organization
system. For example, requirements mapped to a controlled vocabulary may allow
analyses on the consistency of requirements. Analyses that are targeted at
understanding the completeness of requirements specifications may require
ontologies that are able to reflect structure and relationships between entities.

\subsection{The CC Recommender (CCR)}\label{sec:ccr}
CoClass\footnote{\url{https://coclass.byggtjanst.se/about\#about-coclass}} (CC)
is a taxonomy that describes objects in the construction domain. For our
approach, we make the assumption that the most information-bearing language
construct is the noun. Therefore, to establish traces between requirements and
taxonomy objects, it would make sense to base those traces on nouns. We use a
basic natural language processing pipeline that consists of a segmenter,
tokenizer, stemmer and part-of-speech tagger to identify nouns, using the DKPro
framework~\cite{eckart_de_castilho_broad-coverage_2014}. The domain-specific
terminology found in the taxonomy and in the requirements uses agglutination,
i.e. complex terms are built from two or more component morphemes. Therefore, we
also de-compound the identified nouns with
SECOS~\cite{martin_riedl_unsupervised_2016}.

Once the nouns in a requirement are identified, we associate each noun with
$0..n$ taxonomy objects. This association is established by three predictors.
Each predictor score contributes to a
confidence score $[0..1]$ that is used to rank the taxonomy objects. This strategy allows
us to add new predictors in the future and to weigh the components
contributing to the total score. Next, we describe the currently implemented
predictors that are calculated for each noun found in a requirement.
\paragraph{Exact match predictor} If a stemmed, de-compound noun is found in the requirement and
in the CC taxonomy, the score is computed as follows:
\[P_{exact}=\frac{1}{\mathit{f}_{noun}}\]
where $\mathit{f}_{noun}$ refers to the number of taxonomy objects in which the noun appears in.
The more prevalent the noun is, i.e. the less distinguishing power between objects
it has, the lower the predictive score. 

\paragraph{Semantic similarity predictor} 
While the requirements are written by domain experts, they are not necessarily
using the exact terminology that is used in the taxonomy. We developed therefore
a predictor using word embeddings~\cite{mikolov_efficient_2013}that exploits
semantic relatedness among nouns. Instead of using a pre-trained model,
e.g. from Wikipedia articles, we trained our own domain-specific model. First,
we constructed a text corpus by searching the web
programmatically\footnote{\url{https://azure.microsoft.com/en-us/services/cognitive-services/bing-web-search-api/}}
for nouns used in the labels of CC taxonomy objects. This
resulted in 540,409 documents from which we
extracted\footnote{https://textract.readthedocs.io} the text to construct a
word2vec model\footnote{https://radimrehurek.com/gensim/models/word2vec.html}.
Then, for each noun in a requirement, we use the model to find the 10 most
similar nouns, i.e. ``proxies'', and try to find them in the set of taxonomy nouns. Any such
identified ``proxy'' produces another association between a requirement
noun and a taxonomy object, with the score:

\[P_{similarity}=\frac{1}{\mathit{f}_{proxy} * \cos(\theta_{noun-proxy})}\]

where $\mathit{f}_{proxy}$ refers to the number of taxonomy objects in which the
``proxy'' appears and $\cos(\theta_{noun-proxy})$ refers to the cosine
similarity of noun and ``proxy'' based on the custom word2vec model.
The more similar a ``proxy'' is to a noun found in a requirement, and the
less frequently it appears in the taxonomy, the higher the semantic similarity
predictor score.

\paragraph{History predictor}
Finally, we take into consideration past decisions, that is, data
reflecting whether an association between a particular noun and taxonomy object
was accepted or rejected by the user of the recommender. After a particular
noun-object association has been rejected $n$ times (the default is five, but
can be configured to any number), the predictor score is set
to $-\infty$. Otherwise, the score is calculated as:


\[P_{history}=\frac{\frac{\mathit{f}_{assoc} - min(\mathit{f}_{assoc})}{max(\mathit{f}_{assoc}) - min(\mathit{f}_{assoc})}}{\mathit{f}_{noun}}\]

where $\mathit{f}_{assoc}$ refers to the number of existing associations between the
noun and taxonomy object. In the numerator of the fraction we scale the
frequency of occurrences to $(0..1]$. All predictors produce scores in this
range, which allows us to calculate an overall confidence score:

\[P_{confidence} = \frac{P_{exact} + P_{similarity} + P_{history}}{3} \]

We have implemented $CCR$ using INCEpTION~\cite{klie_inception_2018}, a web-based
annotation platform. The user is presented a requirement and for each
(recognized) noun, one or more suggested associations with taxonomy objects are
shown. The suggestions are ordered by the calculated $P_{confidence}$ score. The
user can then either reject or accept suggestions until no more suggestions
are available. The source code for $CCR$, the instrumentation of the experiment,
and the collected data together with the statistical analysis is available online\footnote{\url{https://zenodo.org/record/3827169}}.

\subsection{Related Work}\label{sec:rw}
We direct readers to the systematic literature review by Dermeval et
al.~\cite{dermeval_applications_2016}, who reviewed the research on ontologies as
knowledge organization systems supporting requirements engineering activities,
and to Borg et al.~\cite{borg_recovering_2014} who reviewed information
retrieval approaches to traceability recovery.

\begin{table*}
  \begin{threeparttable}
    \footnotesize
    \caption{Pre-questionnaire results}\label{tab:quest} \begin{tabular}{p{0.1\textwidth}p{0.1\textwidth}p{0.1\textwidth}p{0.1\textwidth}p{0.1\textwidth}p{0.1\textwidth}p{0.1\textwidth}p{0.1\textwidth}}
      \toprule
      \multicolumn{1}{c}{Variables} & \multicolumn{4}{c}{$CCR$} & \multicolumn{3}{c}{$search$} \\
      \cmidrule(lr){2-5}
      \cmidrule(lr){6-8}
                & P1 & P2 & P3 & P4 & P5 & P6 & P7 \\
      \cmidrule(lr){1-5}
      \cmidrule(lr){6-8}
      Current role & Product owner asset management & Contract specialist & Proj. manager technical requirements & Information management research & Proj. manager technical requirements  & Bridge specialist & Proj. manager technical requirements  \\
      \cmidrule(lr){1-5}
      \cmidrule(lr){6-8}
      Years in role & 5 & 10 & 6 & 2 & 5 & 9 & 6 \\
      \cmidrule(lr){1-5}
      \cmidrule(lr){6-8}
      Total exp. & 5 & 10 & 23 & 15 & 25 & 35 & 23 \\
      \cmidrule(lr){1-5}
      \cmidrule(lr){6-8}
      Writing requirements & once a month & a couple of times per year & a couple of times per year & a couple of times per year & a couple of times per year & never & a couple of times per year \\
      \cmidrule(lr){1-5}
      \cmidrule(lr){6-8}
      Read requirements & once a month & daily & once a week & a couple of times per year & daily & once a week & once a week \\
      \cmidrule(lr){1-5}
      \cmidrule(lr){6-8}
      Experience CC & yes & no & yes & yes & yes & no & no \\
      \cmidrule(lr){1-5}
      \cmidrule(lr){6-8}
      Use of CC & once a month & N/A & a couple of times per year & a couple of times per year & daily & N/A & N/A \\
      \cmidrule(lr){1-5}
      \cmidrule(lr){6-8}
      Location & onsite & offsite & onsite & onsite & onsite & offsite & onsite \\
      \bottomrule
    \end{tabular}
    \begin{tablenotes}
      \item Available frequency options: daily, once a week, once a month, a couple
        of times per year, less frequently, never                                             
    \end{tablenotes}
  \end{threeparttable}
\end{table*}

\section{Experiment Planning}\label{sec:ep}
We have designed a quasi-experiment with industry practitioners, following the
guidelines by Wohlin et al.~\cite{wohlin_experimentation_2000} and report the
experiment planning according to Jedlitschka et
al.~\cite{jedlitschka_reporting_2008}. We decided against a randomized design as
we wanted to balance control and treatment group with respect to the
participants' experience on requirements and the domain-specific taxonomy.
Furthermore, we decided to involve industry practitioners to avoid
constructing artificial domain-specific material (requirements, taxonomy) that
would be suitable for e.g. student subjects. 

\subsection{GQM}\label{sec:gqm}
We refine the research question posed in the introduction with the
Goal-Question-Metric (GQM) approach~\cite{basili_goal_1994}.

\paragraph{Goal} Analyze manual and recommender aided association of
requirements with a taxonomy, for the purpose of evaluation with respect to
efficiency, accuracy, consistency and confidence from the viewpoint of
domain experts in the context of an infrastructure project.

The evaluation aspects of efficiency, accuracy, consistency and reliability were
inspired by work on effectiveness evaluation of expert
systems~\cite{sharda_decision_1988}.

\paragraph{Questions}
\begin{enumerate}
  \item[Q1] Is there a difference in time spent to create manual and
        recommender aided associations?
  \item[Q2] Is there a difference in accuracy between manual and recommender
    aided associations?
  \item[Q3] Is there a difference in consistency between manual and
    recommender aided associations?
  \item[Q4] Is there a difference in reported confidence by engineers
    creating manual and recommender aided associations?
  \end{enumerate}

  \paragraph{Metrics} Table~\ref{tab:metrics} maps metrics to questions.
  \begin{center}
    \footnotesize
    \captionof{table}{Metrics to answer questions}\label{tab:metrics}
    \begin{tabular}{p{0.1\columnwidth}p{0.8\columnwidth}}
      \toprule
      Question & Metrics\\
      \midrule
      Q1 & time spent per requirement (M1)\\
      Q2 & expert-based judgment on correctness of associations (M2)\\
      Q3 & within-group variation of made associations (M3)\\
      Q4 & self-reported confidence in terms of completeness (M4) and correctness (M5) of made associations\\
      \bottomrule
    \end{tabular}
  \end{center}

  We measure M1 by the time in seconds spent to associate a requirement with
  zero or more objects from the taxonomy. The correctness of associations (M2)
  is assessed independently by two domain experts. They are tasked to distribute
  10 points to the instances of associations made by the experiment
  participants. The judgment is made without knowledge of whether the
  associations were made with or without the aid of the recommender. We measure
  thereby the relative and not absolute correctness of the associations. We
  measure variance within the group (M3) by encoding each requirement as a
  vector that represents the associated taxonomy objects. The larger the angle
  between the vectors representing the association instances, the larger the
  within group variation (and the lower the association consistency). The idea
  for this measure stems from the vector space model~\cite{salton_vector_1975}
  that is often used to encode text documents to analyze their similarity.
  Finally, we measure confidence in completeness (M4), i.e. whether all relevant
  taxonomy objects were associated with the requirement, and correctness (M5),
  i.e. whether the made associations are correct. The experiment participants
  self-report their confidence on a scale from -2 to +2 per requirement.

\subsection{Hypotheses and Variables}
Based on the metrics defined in the GQM, we formulate five hypotheses pairs. We
show one pair and explain next how the five pairs are generated.

\begin{equation}\label{h1}
  \begin{split}
    H_{0n}: Mn_{CCR} = Mn_{search}\\
    H_{1n}: Mn_{CCR} \neq Mn_{search}
  \end{split}
\end{equation}

where $n=[1..5]$ and the dependent variables $M_{n=[1..5]}$ refer to the metrics
defined in Table~\ref{tab:metrics}. The factor is how the association between
requirement and CC object is supported in INCEpTION, with two possible
treatments. $CCR$ refers to the treatment using the CC recommender we described
in Section~\ref{sec:ccr}, while $search$ refers to the treatment using the
full-text search to find CC taxonomy objects in INCEpTION.

Note that our hypothesis formulations do not assume directionality because we do
not compare an established and a new method. Both alternatives, associating
requirements to CC objects with and without $CCR$ support, are activities that are not
familiar to the participants (control of experience is discussed in
Section~\ref{sec:pp}).
  
\subsection{Participants}\label{sec:pp}
  We recruited seven domain experts with varying experience working with
  requirements and the domain-specific taxonomy. The participants filled in a
  questionnaire that we used balance treatment and control group, based on
  experience with reading and writing requirements, use of the CC taxonomy and
  their overall experience in the construction domain (Table~\ref{tab:quest}).
  Two remotely participating subjects were equally distributed between treatment
  and control group. There was an
  uneven number of participants and we chose to add four to the $CCR$ treatment
  in order to collect more usage experience on the instrument under
  investigation.

\subsection{Materials}
We randomly sampled 100 from a set of 1,216 requirements that belong to an ongoing
infrastructure project. The average number of words per requirements was 19
(minimum: 5, maximum: 77). The CC taxonomy contained 1,420 objects. Each object
has a label, a description and associated synonyms. Finally, the participants
were given a spreadsheet in which they reported the duration (M1) and confidence
(M4, M5) for each requirement they mapped to objects in the CC taxonomy.

\subsection{Tasks}
The participants used the web-based annotation system INCEpTION.
Both groups were tasked to annotate the same requirements, in the same order. The
only difference between the two groups was that the treatment group received
suggestions from $CCR$, which they either accepted or rejected, while the control group
used the built-in search functionality of the annotation tool to find the
relevant objects.

\subsection{Experiment Design}
We chose for the pilot a simple one factor and two treatments
design~\cite{wohlin_experimentation_2000}. All participants received the same
requirements and were using the assigned treatment throughout the experiment.
We assigned four participants to the $CCR$ and three participants to the
$search$ treatment (see Table~\ref{tab:quest}).

\subsection{Procedure}
We carried the pilot experiment out on January 15, 2020. We allocated one hour
for explaining the principle idea of the activity, associating requirements to
CC objects, and illustrated the mechanics of the annotation task with INCEpTION,
both using the $CCR$ and the built-in search functionality. The remainder of the
time (two hours) was allocated to perform the experimental task. All
participants, except two who were connected via a videoconferencing software,
were located in the same room. While the participants performed the tasks, the
author of this paper answered questions and helped participants in case of
technical issues.

\subsection{Analysis Procedure}
Due to the low number of observations collected in the pilot experiment, we
resolved to a non-parametric test statistic with lower power
(Mann-Whitney-Wilcoxon) rather than its parametric counterpart (t-test).
While the results are therefore less robust, we deem it
important to also pilot the analysis procedure, especially since the raw data
required intermediate analyses to evaluate accuracy
(Section~\ref{sec:acc}) and consistency (Section~\ref{sec:cons}), as described
in the respective sections.

\subsection{Deviations from the Plan}
We spent 1.75 hours, instead of the allocated 1 hour, on the experiment
introduction. Including a 15 minute break, 1 hour was spent on the
experiment execution, instead of the originally planned 2 hours.

\section{Analysis}\label{sec:ana}
During the allocated time for the experiment execution, the participants completed
a varying number of tasks (requirements), shown in Table~\ref{tab:taskscompleted}. We limit
therefore our analysis to the requirements that were annotated by all
participants ($n_{requirements}=7$). Next, we analyse the results, answering our
four GQM questions.

\begin{center}
  \footnotesize
  \captionof{table}{Finished tasks by participants P1-P7}\label{tab:taskscompleted}
    \begin{tabular}{llllllll}
      \toprule
      & \multicolumn{4}{c}{$CCR$} & \multicolumn{3}{c}{$search$} \\
      \cmidrule(lr){2-5}
      \cmidrule(lr){6-8}
      & P1 & P2 & P3 & P4 & P5 & P6 & P7 \\
      \cmidrule(lr){2-5}
      \cmidrule(lr){6-8}
      Number of tasks & 28 & 32 & 24 & 14 & 12 & 13 & 7 \\
      Median time per task (s) & 72 & 59 & 69 & 179 & 105 & 17 & 128  \\
      \bottomrule
    \end{tabular}
  \end{center}
    
  \subsection{Efficiency}
  Median duration in groups $CCR$ and $search$ was 61 and 101 seconds. The
  distribution in the two groups did not differ significantly
  (Mann-Whitney-Wilcoxon $U = 209$, $n_{CCR}=28$, $n_{search}=21$, $p=0.09$).
  We cannot reject $H_{01}$ at $\alpha < 0.05$. Figure~\ref{fig:duration}
  shows a box plot of the annotation duration for each requirement. 
\begin{center}
  \includegraphics[width=\columnwidth]{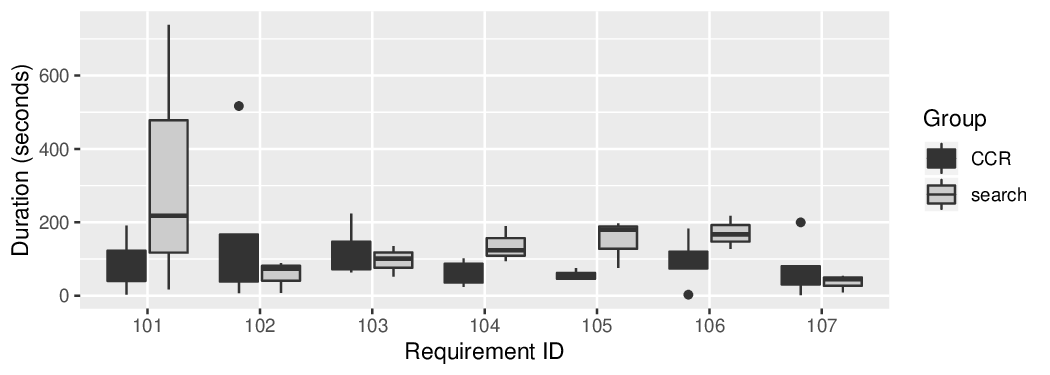}
  \captionof{figure}{Annotation duration (seconds) per requirement}\label{fig:duration}
\end{center}

\subsection{Accuracy}\label{sec:acc}
The participants produced 49 observations, i.e. associations between requirements and CC
objects. Two domain experts evaluated their relative accuracy (see
Section~\ref{sec:gqm}) by distributing 10 points on the associations for each
requirement. Table~\ref{tab:agreement} shows the frequency of the evaluators
agreements/disagreements and the score they were apart, indicating that the two
experts had a good agreement.
\begin{center}
  \footnotesize
  \captionof{table}{Inter-rater agreement}\label{tab:agreement}
    \begin{tabular}{ll}
      \toprule
      Agreements (score difference = 0) & 23 \\
      Disagreements (score difference $=$ 1) & 17 \\
      Disagreements (score difference $=$ 2) & 6 \\
      Disagreements (score difference $=$ 3) & 3 \\
      Disagreements (score difference $>$ 3) & 0 \\
      \bottomrule
    \end{tabular}
  \end{center}
The median accuracy score in groups $CCR$ and $search$ was 4 and 8. The
distribution in the two groups differed significantly (Mann-Whitney-Wilcoxon $U
= 0$, $n_{CCR}=n_{search}=7$, $p = 0.002$).
We reject $H_{02}$ at $\alpha < 0.05$. Looking at the bar
plot in Figure~\ref{fig:accuracy}, which shows the average score of the two
evaluators, we see that $search$ resulted in more accurate results than $CCR$. 
\begin{center}
  \includegraphics[width=\columnwidth]{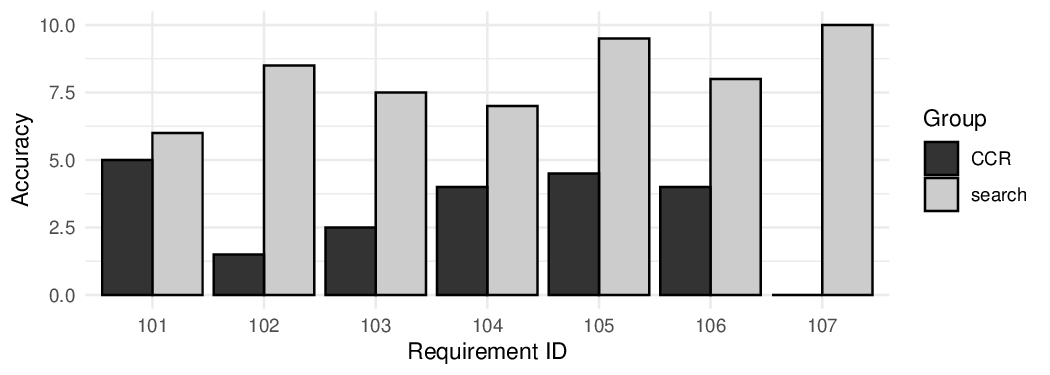}
  \captionof{figure}{Association accuracy per requirement}\label{fig:accuracy}
\end{center}

\subsection{Consistency}\label{sec:cons}
Table~\ref{tab:consistency} shows an example of how four participants annotated a
requirement. Each term $T_{1..10}$ is coded with a label representing a CC object (1
representing no object). For example, participant P1 associated term T8 with object 2 while
P2, P3 and P4 associated the same term with object 5.  
\begin{table}
  \centering
  \footnotesize
  \caption{Example of coding to assess consistency}\label{tab:consistency}
    \begin{tabular}{lllllllllll}
      \toprule
      & T1 & T2 & T3 & T4 & T5 & T6 & T7 & T8 & T9 & T10\\
      \midrule
      P1 & 1 &	1 &	1 &	1 &	1 &	1 &	1 &	2 &	1 &	3\\
      P2 &1	& 1 &	1	& 1 &	1 & 1 &	1 &	5 &	1 &	1\\
      P3 & 1	& 1	& 1	& 1	& 1	& 1	& 1	& 5	& 1	& 1\\
      P4 & 1	& 1	& 1	& 1	& 1	& 1	& 1	& 5	& 1	& 3\\
      \bottomrule
    \end{tabular}
  \end{table}
In order to assess consistency between participants within one treatment
group, we calculated the average pairwise cosine similarity between their
association vectors which results in a score between $(0..1]$, 1 indicating
complete consistency.
The median consistency score in groups $CCR$ and $search$ was 0.98 and 0.90. The
distribution in the two groups did not differ significantly
(Mann-Whitney-Wilcoxon $U=34$, $n_{CCR}=n_{search}=7$, $p = 0.25$).
We cannot reject $H_{03}$ at $\alpha < 0.05$.
Figure~\ref{fig:consistency} illustrates the results.

\begin{center}
  \includegraphics[width=\columnwidth]{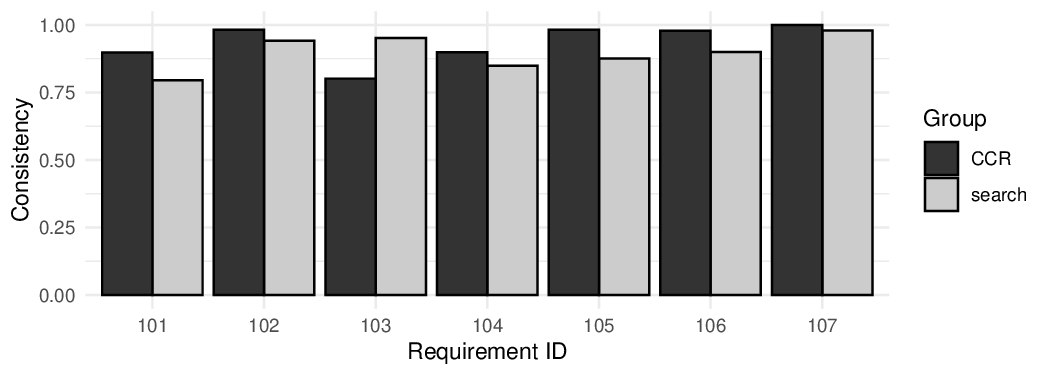}
  \captionof{figure}{Consistency per requirement}\label{fig:consistency}
\end{center}

\subsection{Confidence}
Figures~\ref{fig:correctness} and \ref{fig:completeness} show the results of the
participants' self-reported confidence ($[-2,+2]$) in terms of correct and
complete associations per requirement.

\begin{center}
  \includegraphics[width=\columnwidth]{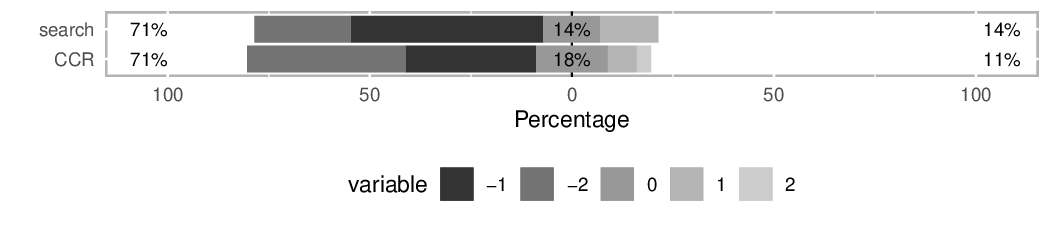}
  \captionof{figure}{Correctness confidence}\label{fig:correctness}
\end{center}

\begin{center}
  \includegraphics[width=\columnwidth]{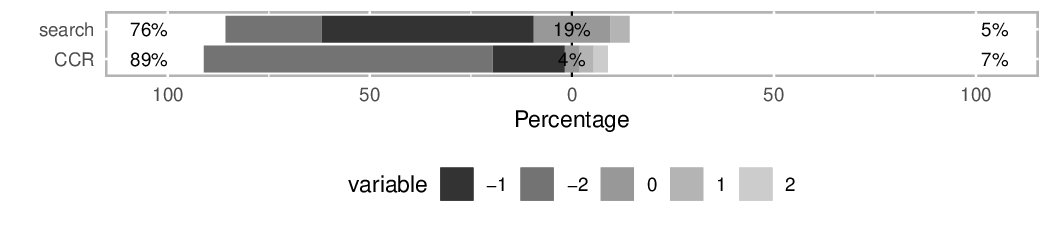}
  \captionof{figure}{Completeness confidence}\label{fig:completeness}
\end{center}

Correctness confidence in groups $CCR$ and $search$ is low (71\%), neutral (18\%
and 14\%) and high (11\% and 14\%). The distribution in the two groups did not
differ significantly (Mann-Whitney-Wilcoxon $U=260$, $n_{CCR}=28$, $n_{search}=21$, $p=0.47$).
We cannot reject $H_{04}$ at $\alpha < 0.05$.

Completeness confidence in groups $CCR$ and $search$ is low (89\% and 76\%),
neutral (4\% and 19\%) and high (7\% and 5\%). The distribution in the two
groups did differ significantly (Mann-Whitney-Wilcoxon $U=162$, $n_{CCR}=28$,
$n_{search}=21$, $p=0.004$).
We reject $H_{05}$ at $\alpha < 0.05$, and accept that the self-reported
confidence in terms of completeness is higher with $search$.

\section{Discussion}\label{sec:dis}
The feedback collected during the experiment and the results on the
self-reported confidence on the correctness and completeness of the associations
between requirements and CC taxonomy objects indicate that the task is
challenging, even for engineers with extensive domain experience. Both
approaches, $CCR$ and $search$ lead to low confidence on the created traces.
While there are indications that traces can be created faster with $CCR$,
$search$ traces were judged as more accurate. We received feedback during the
experiment that the $CCR$ suggested associations do not consider the context of
the requirement. This corresponds well with the implementation of the used
predictors which consider currently only single nouns.

\subsection{Threats to Validity}
We limit our discussion to two major threat dimensions~\cite{wohlin_experimentation_2000}.
\paragraph{Conclusion} The concept of tracing a requirement to a taxonomy
was new to all participants. The training was very limited and the
participants may have interpreted the task in different ways.
\paragraph{Internal} The annotation UI within INCEpTION, where suggestions are
accepted with a single click and rejected with a double click, caused some
confusion and the participants perceived it as error prone. Furthermore, the manual collection
of spent time and confidence is error prone. Finally, the participants may have
influenced each others answers by working in the same room, on the same
requirements at the same time.

\section{Conclusions}\label{sec:con}
We proposed early requirements tracing to domain-specific taxonomies to support
the analysis of requirements specifications. We developed a recommender that
suggests associations between requirements and a taxonomy in the
construction domain. To evaluate the feasibility of creating such traces,
we designed a controlled experiment, comparing the recommender with manually
establishing associations. We measured multiple dimensions to better understand
the differences between the approaches. As a pilot, the experiment was
successful since it provided initial feedback on the performance of the
recommender, insight on the experimental material and illustrated that the collected
data can be meaningfully analysed. Future experiments can also consider the
factors participant experience and the length of requirements. Furthermore, the
idea of using a taxonomy as a mediator to establish trace links needs to be
further validated on other artefacts than requirements, such a design
documentation or source code.

\section*{Acknowledgments}
The author would like to thank Claes Wohlin for providing feedback on the
experiment design. This work was funded by Trafikverket (FoI KREDA).

\bibliographystyle{IEEEtran}
\bibliography{main}

\end{document}